\documentclass[epj]{svjour}

\usepackage{graphicx}
\begin{document}
\title{The BMV experiment : a novel apparatus to study the propagation of light in a transverse magnetic field}
\author{R\'emy Battesti\inst{1}, Beno\^it Pinto Da Souza\inst{2}, S\'ebastien Batut\inst{1}, C\'ecile Robilliard\inst{2},
%Mathilde Fouch\'e\inst{1},
Gilles Bailly\inst{2},
%Julien
%Mauchain\inst{2},
Christophe
Michel\inst{3}, Marc Nardone\inst{1}, Laurent Pinard\inst{3},
Oliver Portugall\inst{1}, Gerard Tr\'enec\inst{2}, Jean-Marie Mackowski\inst{3}, Geert L.J.A. Rikken\inst{1}, Jacques Vigu\'e\inst{2} \and Carlo Rizzo\inst{2}% etc
}                     % Do not remove
\offprints{carlo.rizzo@irsamc.ups-tlse.fr}          % Insert a name or remove this line
\institute{Laboratoire National des Champs Magn\'{e}tiques
Puls\'{e}s (UMR 5147, CNRS - INSA - Universit\'{e} Paul Sabatier
Toulouse 3), 31400 Toulouse cedex, France. \and Laboratoire
Collisions Agr\'{e}gats R\'{e}activit\'{e} (UMR 5589, CNRS -
Universit\'{e} Paul Sabatier Toulouse 3), IRSAMC, 31062 Toulouse
Cedex 9, France. \and Laboratoire des Mat\'{e}riaux Avanc\'{e}s
(LMA) (Universit\'{e} Claude Bernard Lyon 1, CNRS, IN2P3),
Villeurbanne Cedex, France}
\date{Received: date / Revised version: date}
\abstract{ In this paper, we describe in detail the BMV
(Bir\'efringence Magn\'etique du Vide) experiment, a novel
apparatus to study the propagation of light in a transverse
magnetic field. It is based on a very high finesse Fabry-Perot
cavity and on pulsed magnets specially designed for this purpose.
We justify our technical choices and we present the current status
and perspectives.
\PACS{
      {PACS-key}{discribing text of that key}   \and
      {PACS-key}{discribing text of that key}
     } % end of PACS codes
} %end of abstract
\maketitle
\section{Introduction}
\label{intro} Experiments on the propagation of light in a
transverse magnetic field date from the beginning of the 20th
century. In 1901 Kerr \cite{Kerr} and in 1902 Majorana
\cite{Majorana} discovered that linearly polarized light,
propagating in a medium in the presence of a transverse magnetic
field, acquires an ellipticity. In the following years, this
magnetic birefringence has been studied in details by A. Cotton
and H. Mouton \cite{CottonMouton} and it is known nowadays as the
Cotton-Mouton effect. The velocity of light propagating in the
presence of a transverse magnetic field B depends on the
polarization of light, i.e. the index of refraction $n_\parallel$
for light polarized parallel to the magnetic field is different
from the index of refraction $n_\perp$ for light polarized
perpendicular to the magnetic field. For symmetry reasons, the
difference  $\Delta n = (n_\parallel - n_\perp)$ is proportional
to $B^2$. Thus, in general an incident linearly polarized light
beam exits elliptically polarized from the magnetic field region.
The ellipticity to be measured $\Psi$ can be written as

\begin{equation}\label{Elli}
    \Psi = \pi \frac{L}{\lambda} \Delta n \sin{2\theta}
\end{equation}

where $L$ is the optical path in the magnetic field region,
$\lambda$ the wavelength of the light, and $\theta$ the angle
between light polarization and the magnetic field.

In dilute matter like gases, such an effect is usually very small
and it needs very sensitive ellipsometers to be measured. {\it Ab
initio} calculations can be performed using the most advanced
computational techniques and they still remain very challenging
\cite{RizzoRizzo}.

Propagation of light in vacuum in the presence of a transverse
magnetic field has been experimentally studied since 1929
\cite{Watson}. The first motivation was to look for a magnetic
moment of the photon. Only around 1970, thanks to the effective
Lagrangian established in 1935 and 1936 by Kochel, Euler and
Heisenberg \cite{EulerKochel} \cite{HeisenbergEuler}, it has been
shown that the Cotton-Mouton effect should also exist in a vacuum
\cite{Bialynicka-Birula} \cite{Adler}. Quantum ElectroDynamics
(QED) predicts that a field of 1 T should induce an anisotropy of
the index of refraction of about $4\times10^{-24}$. This very
fundamental prediction has not yet been experimentally verified.

Some of the earlier experiments were based on the use of an
interferometer of the Michelson-Morley type. One of the two arms
passed through a region where a transverse magnetic field was
present inducing a difference in the light velocity that should
have been observed as a phase shift \cite{FarrBanwell}
\cite{BanwellFarr}. In 1979, Iacopini and Zavattini
\cite{IacopiniZavattini} proposed to measure the ellipticity
induced on a linearly polarized laser beam by the presence of a
transverse magnetic field using an optical cavity in order to
increase the optical path in the field. The effect to be measured
was modulated in order to be able to use heterodyne technique to
increase the signal to noise ratio.

In 1986, Maiani, Petronzio, and Zavattini \cite{Maiani} showed
that hypothetical low mass, neutral, spinless boson, scalar or
pseudoscalar, that couples with two photons could induce an
ellipticity signal in the Zavattini apparatus similar to the one
predicted by QED. Moreover, an apparent rotation of the
polarization vector of the light could be observed because of
conversion of photons into real bosons resulting in a vacuum
magnetic dichroism which is absent in the framework of standard
QED. The measurements of ellipticity and dichroism, including
their signs, can in principle completely characterize the
hypothetical boson, its mass $m_a$, the inverse coupling constant
$M_{a}$, and the pseudoscalar or scalar nature of the particle.
Maiani, Petronzio, Zavattini's paper was essentially motivated by
the search for Peccei and Quinn's axions. These are pseudoscalar,
neutral, spinless bosons introduced to solve what is called the
{\it strong CP problem} \cite{Peccei}. However, it was soon clear
that such an optical apparatus could hardly exclude a range of
axion parameters not already excluded by astrophysical bounds
\cite{Raffelt}.

Following Zavattini's proposal, after tests at CERN in Geneva,
Switzerland, \cite{TestCERN}, an apparatus has been set up at the
Brookhaven National Laboratory, USA \cite{Cameron}. No evidence
for dichroism or for ellipticity induced by the magnetic field was
found. The sensitivity being insufficient to detect the QED
effect, only limits on the axion parameters could be published in
1993 \cite{Cameron}, $M_{a}
> 2.8\times10^{6}$ GeV at the 95\% confidence level,
provided $m_a < 6.7\times10^{-3}$ eV.

In 1991, a new attempt to measure the vacuum magnetic
birefringence has been started at the LNL in Legnaro, Italy, by
the PVLAS collaboration \cite{Bakalov}. This experiment is again
based on ref. \cite{IacopiniZavattini}. A Fabry-Perot cavity is
used to increase the effect to be measured, while a
superconductive 5 T magnet rotates around its own axis to modulate
it. Eventually, the collaboration has published the observation of
a magnetically induced dichroism in vacuum \cite{Zavattini}, and
also of magnetically induced ellipticity in excess of what is
expected according to QED \cite{Karuza}.  These results have
triggered a lot of interest in the field, in particular because
the existence of axions could be the explanation for this
unexpected signal \cite{Lamoreaux}. The proposed range is
$2\times10^{5}$ GeV $<$ $M_{a}$ $< 5.9\times10^{5}$ GeV, provided
that $1\times10^{-3}$ eV $< m_a < 1.5\times10^{-3}$ eV. Recently,
the PVLAS collaboration has disclaimed their previous
observations. An ellipticity signal is still present at 5.5 T,
while no ellipticity signal is observed at 2.3 T \cite{Disclaim}.
No clear explanation is given as to why the original signal has
disappeared.

As soon as the original PVLAS results have become known, they have
created big expectations, and the research field has gained a
large momentum since this could be the first positive sighting of
axions or axionlike particles \cite{Lamoreaux}.  Indeed, the PVLAS
result is in contradiction with the solar axion search, and in
particular with the results of the CERN axion telescope CAST
\cite{CAST}, which exclude axionlike particles with a mass smaller
than $10^{-2}$ eV and a inverse coupling constant smaller than
$10^{10}$ GeV. To accommodate both results one should change the
nature of axionlike particle interaction to justify their
confinement to the interior of the sun (see e.g.
\cite{MassoRedondo}).

From the experimental point of view, different proposals have been
put forward (see e.g. \cite{iasedu}), most of them planning
photoregeneration experiments.  Such a kind of experiment is based
on the idea that once an axionlike particle is created by photon
conversion in a magnetic field, the particle escapes from the
magnet region while the light beam can be easily confined. Now, if
the created particle passes through another magnet region, it can
be converted back into a photon that one can detect in a region
where no photon should exist \cite{VanBibber}. QED vacuum
birefringence is also put forward in the proposal OSQAR
\cite{OSQAR}. Most of these projects are expected to take data in
2007. First results of a photoregeneration experiment specially
designed to test the PVLAS claim have appeared very recently
\cite{Robilliard} excluding the axionlike particle interpretation
of the PVLAS original result with a confidence level greater than
99.9\%.

In the meantime new results have also been posted on the arXiv web
site by the Q\&A (Quantum electrodynamics test \& search for
Axion) project based at National Tsing Hua University, Taiwan.
This project has started around 1996. The experimental set-up is
similar to the PVLAS one. No effect has been detected, in
contradiction with the original result from the PVLAS
collaboration. As far as we understand, the confidence level is
about 68\% \cite{Taiwan}.

In this paper, we describe a novel apparatus to study the
propagation of light in a transverse magnetic field : the BMV
(Bir\'efringence Magn\'etique du Vide) experiment. It is based on
a very high finesse Fabry-Perot cavity and on a pulsed transverse
magnet specially designed for this purpose. We justify our
technical choices and we present the current status and
perspectives. We show that our apparatus has the sensitivity to
test the PVLAS result, and eventually to measure the vacuum
magnetic birefringence. A very important QED prediction that has
been waiting to be tested since about 75 years.

\section{The optical apparatus}
\label{sec:3}

\begin{figure}[h]
  \includegraphics[width=9cm]{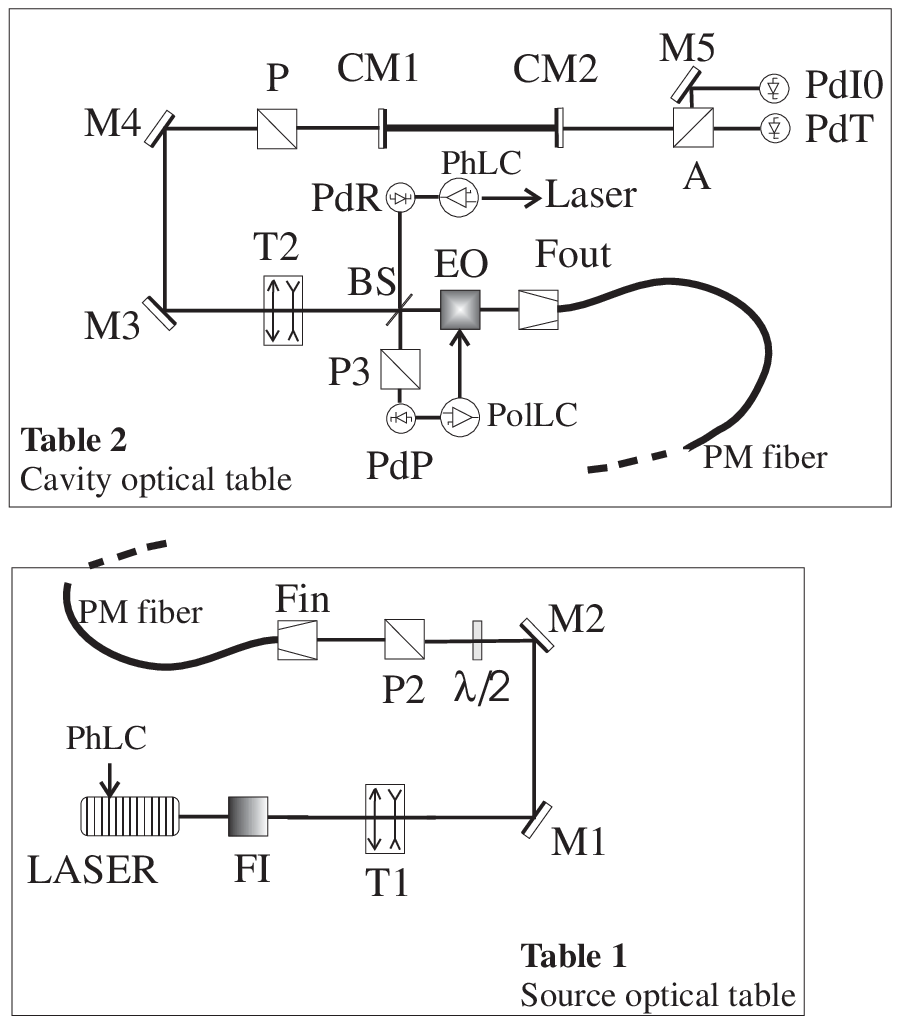}\\
  \caption{Scheme of the optical apparatus.}\label{Sch}
\end{figure}

In fig. \ref{Sch}, we present a sketch of our optical apparatus.
The optical cavity and the laser source are put on two different
optical tables, linked by a polarization maintaining optical fiber
(PM fiber).

On the first table, light emitted by a mono-mode frequency
controlled Nd:YAG laser at a wavelenght $\lambda=1064$ nm passes
through a Faraday isolator FI.
%and .

A telescope T1 adapts the beam waist to be injected into the
fiber. Mirrors M1 and M2 transport the beam to a $\lambda/2$
waveplate and a polarizer prism P2 which allow to adjust beam
intensity at the fiber entrance. Fiber collimator Fin launches
light into the fiber with a coupling efficiency of about 85 $\%$.

Light exits by the collimator Fout on the second table and it
passes through an electro-optical modulator EO. This modulator is
part of a feedback loop designed to lock the beam polarization.
This loop is constituted by a beamsplitter BS that sends a
fraction of light to a polarizer prism P3, a photodiode PdP
detects the light transmitted by P3. The signal of PdP drives a
locking circuit PolLC that maximizes the transmission through P3
by acting on the EO modulator.

Light going through the beamsplitter BS is matched to the cavity
waist by a telescope T2 and transported by mirrors M3 and M4.
These two mirrors are also used to align the optical beam with the
optical axis of the Fabry-Perot cavity made by two very high
reflectivity mirrors CM1 and CM2.

A system of high precision translations and rotations for the
Fabry-Perot cavity mirrors and also for the polarizers has been
designed and assembled. The piezoelectric stacks of the mirror
alignment system as well as the mechanism allowing their rotation
are adapted to ultra-high vacuum.

Before entering the Fabry-Perot cavity light is polarized by the
polarizer P. Polarization of the light transmitted by the cavity
is analyzed by the analyzer A. The extraordinary ray is detected
by photodiode PdT, while the ordinary ray after being reflected by
a mirror M5 is detected by photodiode PdI0. Both signals are used
in the data analysis. Mirror M5 turns with prism A and extraction
of the ordinary ray can be done at any position of A.

All the optical components from the polarizer P to the analyzer A
are placed in a ultra high vacuum chamber not shown in the figure.

Light reflected back by the Fabry-Perot cavity arrives at the
beamsplitter BS and it is sent to the photodiode PdR, the signal
of which is used to drive the phase locking circuit PhLC that
locks the laser frequency to the Fabry-Perot cavity.

\subsection{The Fabry-Perot Cavity}

Our Fabry-Perot cavity is made by two identical mirrors of 25.4 mm
(1 inch) diameter. The length of the cavity D in the final version
of our setup should not exceed 2.5 meters. The radius of curvature
C of the mirrors has been chosen equal to 8 m. The resonant
frequencies of the cavity modesis given by the following formula

\begin{equation}\label{ModeFreq}
\nu_{qlm}=\left[q+(l+m+1)\frac{\arccos\left(1-\frac{D}{C}\right)}{\pi}\right]\frac{c}{2D}
\end{equation}

We want that only one mode resonates at the frequency of the
$TEM_{00}$ mode, thus

\begin{equation}\label{ModeFreq00}
\frac{\arccos\left(1-\frac{D}{C}\right)}{\pi}\neq\frac{p}{k+1}
\end{equation}

where $p$ and $k$ are integers. Choosing $C=8$ m guarantees that
only modes with $l+m \gg 1$ can not fulfill eq. \ref{ModeFreq00}.

The minimum waist $w_0$ is situated at the center of the cavity
and it is given by eq. \ref{waist0}.

\begin{equation}\label{waist0}
w_0=\sqrt{\left(\frac{\lambda}{2\pi}\right)\sqrt{D(2C-D)}}
\end{equation}

On the mirror the waist $w_m$ is given by the following formula :

\begin{equation}\label{waistMirror}
w_m=\sqrt{\left(\frac{\lambda}{2\pi}\right)\sqrt{\frac{4C^2D}{2C-D}}}
\end{equation}

For $C=8$ m and $D=2.5$ m, we get  $w_0\simeq$ 1 mm and
$w_m\simeq1.02$ mm.

Assuming a good mode matching, cavity transmission depends on the
mirror transmissivity $T$ and mirror losses $P$. The ratio $r_I$
between the intensity of the transmitted light $I_t$ and the
intensity of the incident light can be written as

\begin{equation}\label{Transm}
r_I=\frac{I_t}{I_i}= \left(\frac{T}{T+P}\right)^2
\end{equation}

The value of the sum $T+P$ is fixed by the value of the requested
finesse since $F=\frac{\pi}{1-R}$ where the mirror reflectivity
$R$ is equal to $1-(T+P)$. For having a finesse between 500 000
and 1 000 000, let's assume that $P=3\times 10^{-6}$ and $T=2
\times 10^{-6}$. In this case $r_I\simeq0.16$.

Let's assume in the following that we want a shot noise limited
sensitivity $\Psi_s$ of the order of $10^{-9}$ rad Hz$^{-1/2}$. As
we show in the following (see eq. \ref{senshomo}), assuming a
quantum efficiency of the photodiode $q=0.7$, one finds that $I_t$
has to be about 40 mW. If $r_I\simeq0.12$, the incidence intensity
$I_i$ has to be about 250 mW.

Now, the intensity $I_m$ on the surface of the exit mirror of the
cavity is $I_t/T$ i.e. about 20 kW. This corresponds to a power
density of the order of 600 kW/cm$^2$.

Mirrors are always unintentionally slightly birefringent (see e.g.
\cite{Brandi}) and therefore the Fabry-Perot cavity resonance line
will be separated in two. Light polarized as the fast axis of
birefringence will see a different optical path with respect to
the light polarized as the slow axis of birefringence. In ref.
\cite{GZavattini} the issue of a birefringent Fabry-Perot cavity
in the framework of ellipticity measurements has been discussed in
detail. We just recall here that the frequency separation of the
two polarizations is given by

\begin{equation}\label{birlin}
\Delta\nu_\delta=\frac{\delta}{2\pi}\frac{c}{2D}
\end{equation}

where $\delta$ is the mirror birefringence. The line width is

\begin{equation}\label{lin}
\Delta\nu=\frac{1}{F}\frac{c}{2D}
\end{equation}

To avoid that the cavity acts as polarizer prism, it is necessary
that $\Delta\nu_\delta < \Delta\nu$, thus

\begin{equation}\label{delta}
\delta < \frac{2\pi}{F}
\end{equation}

Thus for a 1 000 000 finesse cavity $\delta < 6.28\times10^{-6}$,
which is demanding but which has already been observed for high
finesse mirrors \cite{Brandi}.

\subsubsection{Magnetic field modulation and cavity finesse}

In our project, we plan to use a very high finesse cavity to
increase the ellipticity signal to be measured. For 1 000 000
cavity finesse and 2 meters cavity length, the photon lifetime  in
the cavity $\tau = F D/(\pi c)$ is about 2 msec. This duration is
not negligible with respect to the magnetic pulse duration.

In the following, we calculate the output field of a Fabry-Perot
interferometer in the case where the optical path is a slowly
varying function of time. More precisely, the phase acquired by
the light wave while going forward and backward in the Fabry-Perot
cavity is given by:

\begin{equation}
\label{a1} \psi = 2 k D + \phi(t)
\end{equation}

\noindent We assume that $\phi(t)$ has negligible variations over
the time $ t_D = D/c$ taken by the light wave to go from one
mirror to the other. However, because the cavity has a very high
finesse $F$, the variations of $\phi(t)$ over the photon lifetime
in the cavity $\tau$ are not necessarily negligible and the goal
of the calculation is to express how the photon lifetime averages
the phase $\phi(t)$. It seems that there is no general solution
for the field exiting from a Fabry-Perot interferometer which is
not stationary in time. The present calculation is approximate and
assumes that the phase $\phi(t)$ is very small ($\phi(t)$ could be
large but the varying part must be small and we assume that there
is no time independent part).

We calculate the light field exiting from the Fabry-Perot at the
instant $t$, by summing the contributions of the rays which have
been transmitted with $n$ return paths:

\begin{center}
\begin{eqnarray}
\label{a2} \frac{E_{out}(t)}{T E_{in}\exp\left[i
\phi(t-(t_D))\right]}\hspace{0.5cm} =  \\
\nonumber \sum_{j=0}^{\infty} R^j\exp\left[i\left(2jkD +
\sum_{p=1}^j \phi(t-2 p t_D)\right)\right]
\end{eqnarray}
\end{center}

\noindent We assume that the amplitude reflection (respectively
transmission) coefficients of the two mirrors are real and we note
their product as $R$ (respectively $T$). The incident field
$E_{in}$ is taken as perfectly monochromatic. The light which has
been reflected $n$ times at both ends of the Fabry-Perot
interferometer has sampled the phase shift $\phi(t)$ at the times
$(t-2 p t_D)$ with $p$ varying from $1$ to $j$. For the general
case, this expression cannot be evaluated analytically.

We assume that the Fabry-Perot cavity is at resonance ($2kD = 0$
$[2\pi]$) and that the phase $\phi(t)$ is very small so that we
can replace the exponential by its first-order expansion:

\begin{equation}
\label{a3} \exp\left[i \sum_{p=1}^j \phi(t-2 p t_D)\right] \approx
1 + i \sum_{p=1}^j \phi(t-2 p t_D)
\end{equation}

\noindent The condition is not only $\phi(t) \ll 1$ but because
the sum extends over a number of terms comparable to the finesse
$F$, we must have $F \phi(t) \ll 1$. We must calculate two sums:
\begin{eqnarray}
\label{a5} \sum_{j=0}^{\infty} R^j  = \frac{1}{1-R}
\end{eqnarray}

\begin{eqnarray}
\label{a51} \nonumber \sum_{j=1}^{\infty} R^j \sum_{p=1}^j
\phi(t-2 p t_D) &=&\\
\sum_{p=1}^{\infty} \phi(t-2 p t_D)\sum_{j=p}^{\infty} R^j &=& \\
\nonumber \sum_{p=1}^{\infty} \phi(t-2 p t_D) \frac{R^p}{1-R}
\end{eqnarray}

\noindent We can rewrite $R^p$:

\begin{equation}
\label{a6} R^p = \exp\left[- (\Gamma/2) p (2t_D)\right]
\end{equation}
\noindent where $\Gamma = 1/\tau \approx (1-R)/t_D$ and we get:
\begin{eqnarray}
\label{a7} \nonumber \sum_{j=1}^{\infty}R^j \sum_{p=1}^j \phi(t-2
p t_D) \hspace{0.5cm} = \hspace{1cm} \\
\sum_{p=1}^{\infty} \phi(t-2 p t_D) \exp\left[- (\Gamma/2) p
(2t_D)\right] \frac{1}{1-R} \approx   \\
\nonumber  \frac{1}{(1-R)^2} \int_0^{+\infty}(\Gamma/2)
\frac{dt'}{2} \phi(t-t') \exp\left[- (\Gamma/2) t'\right]
\end{eqnarray}

The final result is that the output field is given by:

\begin{equation}
\label{a2} \frac{E_{out}(t)}{T E_{in}\exp\left[i
\phi(t-t_D)\right]} \approx \frac{1}{1-R} \left[ 1 + i \frac
{\langle\phi\rangle}{2(1-R)}\right]
\end{equation}
%\end{eqnarray}

\noindent where
\begin{equation}
\langle\phi\rangle = \int_0^{+\infty}(\Gamma/2){dt'}\phi(t-t')
\exp\left[- (\Gamma/2) t'\right]
\end{equation}

It is easy to make $\phi(t)=0$ in the result and then we find the
traditional result of the Fabry Perot cavity at resonance:

\begin{eqnarray}
\label{a2} \frac{E_{out}(t)}{T E_{in}} = \frac{1}{1-R}
\end{eqnarray}

For the sake of simplicity, let's assume that
$B(t')=B_0\sin(\omega t')$. $B(t')^2$ can be written as

\begin{equation}\label{Bsquare}
B(t')^2=\frac{B_0^2}{4}\left[2-e^{(2i\omega t')}-e^{(-2i\omega
t')}\right]
\end{equation}

\noindent and
\begin{equation}\label{esse}
\langle\phi\rangle \propto \int_0^\infty
\frac{B_0^2}{4}\left[2-e^{(2i\omega (t-\theta))}-e^{(-2i\omega
(t-\theta))}\right]e^{(-(\Gamma/2) \theta)}d\theta
\end{equation}

\noindent where $\theta=t-t'$. Finally, we obtain that

\begin{equation}\label{moy}
\langle\phi\rangle \propto
{B_0^2}\left[1-\frac{1}{\sqrt{1+(\frac{2\omega}{\Gamma/2})^2}}\cos(2\omega
t+\varphi)\right]
\end{equation}

\noindent with $\tan\varphi=-\frac{2\omega}{\Gamma}$

Thus when the magnetic field varies during the lifetime of the
photons in the cavity, the ellipticity acquired by the light
depends on an attenuated averaged value of the square of the
magnetic field and moreover the ellipticity is not in phase with
the square of the field. To avoid such an effect, it is clear that
one needs

\begin{equation}\label{PhiPhi}
\frac{2\omega}{\Gamma/2} \ll 1
\end{equation}

In our experiment, special care will be devoted to minimize such
an effect.

\subsection{Cavity mirrors}
To increase as much as possible our signal to noise ratio, we need
a cavity with a very high finesse. As far as we know, the highest
finesse ever published is about $2\times 10^6$ \cite{Rempe}, while
the highest quality factor $Q$ is the one of the PVLAS cavity
\cite{Zavattini}, $Q\simeq4\times10^{12}$ corresponding to a
cavity linewidth of about 200 Hz and a storage time of 0.5 ms.

\begin{figure}[h]
  \includegraphics[width=9cm]{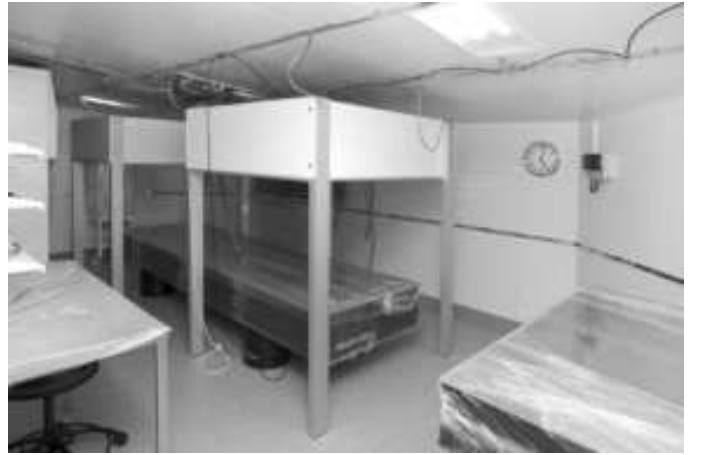}\\
  \caption{Photo of the experimental room.}\label{Salle}
\end{figure}

Our cavity mirrors are made by the Laboratoire des Mat\'{e}riaux
Avanc\'{e}s (LMA). Thanks to their know-how we have currently at
our disposal 4 mirrors with losses ranging from 3 to 9 ppm and
transmission ranging from 1.5 to 1.9 ppm. This corresponds to
finesses ranging from 300 000 to 700 000. The main problem is to
handle these mirrors in an adequate way because they are very
sensitive to pollution. Therefore we have built a special clean
room with laminar flow cabinets for their manipulation. The access
to this room is limited to fully equipped personnel (see fig
\ref{Salle}).

The mirrors are put in a home made kinematic mount which is
activated by two piezoelectric wedges. They allow us to align the
mirror reflecting surface perpendicular to the cavity axis with a
precision better than 4 $\mu rad$. This aligning system operates
in ultra high vacuum. Finally, we use a bronze wheel for rotating
the mirror mount to minimize the cavity birefringence. It allows
us a rotation smaller than 1 mrad.

\subsection{Laser locking to the Fabry-Perot cavity}

The laser source is locked to the Fabry-Perot cavity to maximize
the injection of the light into the magnetic field region using
the usual Pound-Drever-Hall technique \cite{PoundDreverHall}. The
laser crystal itself is used to phase modulate the laser light
\cite{Cantatore}.

The light source is a tuneable non planar ring oscillator Nd:YAG
laser emitting $\sim 200$ mW of power at a wavelength $\lambda =
1064$ nm ($\nu = 2.82\times10^{14}$ Hz). The frequency of the
laser can be changed by two methods: a fast one (bandwidth $>$ 10
kHz) based on a piezoelectric actuator acting on the laser crystal
and a slow one (bandwidth $<$ 1 Hz) based on the change of the
crystal temperature. In practice, the former allows to vary the
laser frequency of about 50$\times10^6$ Hz, whereas the latter
yields changes of about 100$\times10^9$ Hz.

\begin{figure}[h]
  \includegraphics[width=9cm]{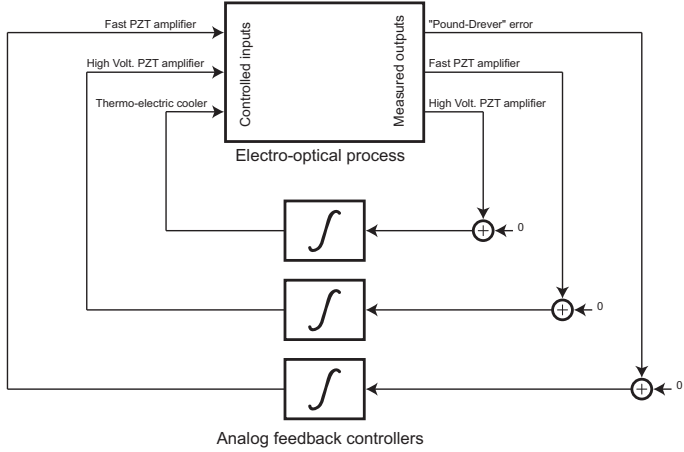}\\
  \caption{Conceptual design of our locking circuit.}\label{Lock}
\end{figure}

In fig. \ref{Lock} we show a simplified scheme of our locking
circuit.

The novelty, compared for example to ref. \cite{Deriva}, is that
we control the laser frequency using three different feedback
signals instead of two. As usual, a very low bandwidth one acts on
the crystal temperature, a second one with an important dynamical
range acts on one of the two ends of the piezoelectric actuator,
and a third one acts on the other end of the piezoelectric
actuator allowing a fine tuning of the laser frequency on the
cavity resonance frequency.

\section{Signal analysis}
\label{sec:1}

\begin{figure}[h]
  \includegraphics[width=7cm]{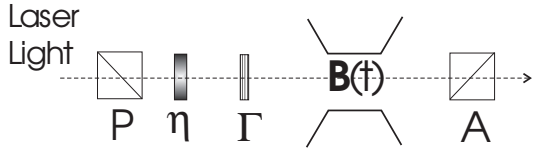}\\
  \caption{Basic principle of the detection technique.}\label{det}
\end{figure}

In fig. \ref{det} we show the basic principle of the apparatus
following ref. \cite{IacopiniZavattini}.

A laser beam is polarized by the polarizer P. Let's assume that
the light intensity after P is $I_t$. Let's also assume that the
light intensity after analyzer A is $I_e$. Going from P to A,
light acquires an ellipticity $\eta(t)$ thanks to an ellipticity
modulator, a static ellipticity $\Gamma$, and an ellipticity to be
measured $\Psi(t)$. In our case $\Psi(t) \propto B(t)^2$. The
detection technique suggested in ref. \cite{IacopiniZavattini} is
the heterodyne one. The same technique has been used in the
experiences of ref. \cite{Cameron} and \cite{Zavattini}.

At the extinction $\sigma^2$, when A is crossed with respect to P,
$I_e$ can be written as

\begin{equation}\label{Ie}
I_e(t)=I_t\sigma^2 + I_t\left[\eta\left(t\right) + \Gamma +
\Psi\left(t\right)\right]^2
\end{equation}

For the sake of simplicity, let us now assume that $\eta(t)=\eta_0
cos(2\pi\Omega_m t + \theta_m)$ and $\Psi(t)=\Psi_0
cos(2\pi\Omega_e t + \theta_e)$, with $\Omega_m \gg \Omega_e$. It
is straightforward to show that $I_e$ is constituted by the
frequency components given in table 1:

\begin{table}
\begin{center}
\begin{tabular}{|c|c|c|c|}
  \hline
  frequency & component & amplitude & phase \\
  \hline
 DC & $I_{DC}$ & {\small $I_t\left(\sigma^2 + \eta_0^2/2 + \Psi_0^2/2 + \Gamma^2\right)$} & 0 \\
 ${\Omega_e}$&  $I_{\Omega_e}$ & $I_t\left(2\Gamma\Psi_0\right)$ & $\theta_e$\\
 ${2\Omega_e}$ & $I_{2\Omega_e}$ & $I_t\left(\Psi_0^2/2\right)$ & $2\theta_e$ \\
 ${\Omega_m}$ & $I_{\Omega_m}$ & $I_t\left(2\Gamma\eta_0\right)$ & $\theta_m$ \\
 ${\Omega_m\pm\Omega_e}$ & $I_{\Omega_m\pm\Omega_e}$ & $I_t\left(\eta_0\Psi_0\right)$ & $\theta_m\pm\theta_e$ \\
 ${2\Omega_m}$ & $I_{2\Omega_m}$ & $I_t\left(\eta_0^2/2\right)$ & $2\theta_m$ \\
  \hline
\end{tabular}
\end{center}
\caption{Frequency components of the signal $I_e$ (see eq.
\ref{Ie})}
\end{table}

The existence of the two $I_{\Omega_m\pm\Omega_e}$ components is
typical of the heterodyne technique in which one beats the effect
to be measured with a carrier effect, $\eta(t)$ in our case.
$I_{\Omega_m\pm\Omega_e}$ components have a linear dependence on
the effect for the detected signal. Let us note that the component
$I_{\Omega_e}$ linear in $\Psi_0$ exists even if $\eta(t)=0$
because of the existence of the spurious static ellipticity
$\Gamma$. Detecting the signal only by modulating the effect is an
example of what is called homodyne technique. Homodyne technique
has the advantage of demanding a simpler optical apparatus
compared to the heterodyne technique because of the absence of the
ellipticity modulator. Homodyne detection is the technique we have
chosen for the BMV experiment. As we show in the following
paragraph, homodyne detection is particularly interesting when
$\Gamma^2 \gg \sigma^2$. This is the case for our BMV experiment.
Actually, $\sigma^2 \approx 10^{-8}$ while $\Gamma^2 \approx
10^{-4}$. $\Gamma$ is due to the birefringence of the mirrors that
constitute our Fabry-Perot cavity.

\subsection{Comparison between heterodyne technique and homodyne technique}

In the following we will justify our choice by comparing the
heterodyne technique and the homodyne one. In particular we
calculate the sensitivity expected using the two techniques. We
prove that they are completely equivalent with respect to this
very important parameter.

In the case of the heterodyne technique one has

\begin{equation}\label{Psi0hete}
\Psi_0=\sqrt{\frac{I_{\Omega_m\pm\Omega_e}^2}{2I_t I_{2\Omega_m}}}
\end{equation}

The homodyne technique, corresponding to $\eta_0=0$, becomes
interesting when $\Gamma^2 \gg \sigma^2, \Psi_0^2/2$. In this case

\begin{equation}\label{Psi0homo}
\Psi_0=\sqrt{\frac{I_{\Omega_e}^2}{4I_t I_{DC}}}
\end{equation}

The limiting noise of both techniques is due to the corpuscular
nature of light (shot noise). Shot noise is proportional to the
square root of the number of photons detected by the photodiode
after the prism A. This is essentially proportional to $I_{DC}$.

\begin{equation}\label{shot}
i_{shot noise}=\sqrt{\frac{2e^2qI_t\left(\sigma^2 + \eta_0^2/2 +
\Psi_0^2/2 + \Gamma^2\right)}{h\nu}}
\end{equation}

\noindent where $e$ is the absolute value of the electron charge,
$q$ is the quantum efficiency  of the photodiode, $h$ is the
Planck's constant, and $\nu$ is the frequency of light. The rate
of photons corresponding to the signal is proportional to
$I_{\Omega_m\pm\Omega_e}$ in the case of heterodyne technique or
to $I_{\Omega_e}$ in the case of the homodyne technique. The
signal-to-noise ratio can be written as

\begin{equation}\label{shothete}
r_{hetero}=\frac{i_{signal}}{i_{shot
noise}}=\sqrt{\frac{2qI_t\Psi_0^2\eta_0^2}{\left(\sigma^2 +
\eta_0^2/2 + \Psi_0^2/2 + \Gamma^2\right)h\nu}}
\end{equation}

\noindent (where the factor 2 in the square root takes into
account the existence of two sidebands) and
\begin{equation}\label{shothomo}
r_{homo}=\frac{i_{signal}}{i_{shot
noise}}=\sqrt{\frac{4qI_t\Psi_0^2\Gamma^2}{\left(\sigma^2 +
\Psi_0^2/2 + \Gamma^2\right)h\nu}}
\end{equation}

\noindent respectively.

It is therefore clear that optimal working conditions thus imply
that $\eta_0^2$ dominates in the $I_{DC}$ component for the
heterodyne technique and that $\Gamma^2$ dominates in the $I_{DC}$
component for the homodyne technique. In this case, let us finally
obtain an expression for the sensitivity $\Psi_{hete}^s$ and
$\Psi_{homo}^s$ respectively. For this we impose the condition of
a signal-to-noise ratio equal to one. We get

\begin{equation}\label{senshete}
\Psi_{hetero}^s=\sqrt{\frac{h\nu}{2qI_t}}
\end{equation}

\begin{equation}\label{senshomo}
\Psi_{homo}^s=\sqrt{\frac{h\nu}{4qI_t}}
\end{equation}

The shot noise limit is hardly ever reached. To get a more general
expression we note that at optimal working conditions equations
\ref{Psi0hete} and \ref{Psi0homo} can also be written as

\begin{equation}\label{Psi0hete2}
\Psi_0=\eta_0\frac{I_{\Omega_m\pm\Omega_e}}{2I_{DC}}
\end{equation}

\begin{equation}\label{Psi0homo2}
\Psi_0=\Gamma\frac{I_{\Omega_e}}{I_{DC}}
\end{equation}

\noindent respectively. In the following, we introduce the
residual ellipticity noise $\gamma(\Omega)$ that is the noise
signal $I_{noise}(\Omega)$ due to the existence of $\Gamma$
divided by $I_{DC}$. This quantity can be measured by looking at
the Fourier spectrum of the signal detected by the photodiode
after the analyzer A set at maximum extinction when $\eta_0=0$ and
no effect is present. We assume that $\eta(t)$ does not add any
extra ellipticity noise. Now, by imposing a signal-to-noise ratio
equal to one, equations \ref{Psi0hete2} and \ref{Psi0homo2} can be
written as

\begin{equation}\label{Psishete}
\Psi_{hete}^s=\eta_0\frac{I_{\Omega_m}
\gamma({\Omega_e})}{2I_{DC}}=\Gamma\frac{2I_{DC}
\gamma({\Omega_e})}{2I_{DC}}=\Gamma\gamma({\Omega_e})
\end{equation}

\begin{equation}\label{Psis0homo}
\Psi_{homo}^s=\Gamma\frac{I_{DC}
\gamma({\Omega_e})}{I_{DC}}=\Gamma\gamma({\Omega_e})
\end{equation}
\noindent respectively, and we find that the heterodyne and
homodyne techniques are completely equivalent. It is also clear
that a sensitivity limited by the shot noise can only be reached
if $\gamma({\Omega_e})$ is small enough. In the case of the
homodyne technique

\begin{equation}\label{gammahomo}
\gamma({\Omega_e}) < \sqrt{\frac{h\nu}{4qI_t\Gamma^2}}
\end{equation}

To obtain a $\Psi_{homo}^s \simeq 10^{-9}$ rad Hz$^{-1/2}$, if
$\Gamma \simeq 10^{-2}$ rad, $\gamma({\Omega_e}) < 10^{-7}$
Hz$^{-1/2}$.

The heterodyne technique needs an external modulation of the
magnetic field and of the ellipticity which is more demanding. In
our experiment we have chosen the homodyne technique ($\eta(t)=0$)
thanks to our pulsed field which gives us a intrinsic modulation.

\subsection{Balanced polarimetry}
A modified version of the apparatus shown in fig. \ref{det} has
also been proposed to measure small ellipticities and in
particular vacuum magnetic birefringence \cite{Rizzo}. A
quarter-wave plate is inserted between the polarizer and the
analyzer. The wave plate is set in such a way that the intensity
of the ordinary and of the extraordinary ray exiting the analyzer
are equal when no birefringence exists along the light path. The
presence of an ellipticity in the beam reaching the analyzer prism
unbalances the intensities of these two rays. The two ray
intensities are measured by two identical photodiodes. The
difference of these two intensities gives the final signal for the
analysis. This detection method is usually called balanced
polarimetry.

One can show that this method is in principle totally equivalent
to the one proposed in ref. \cite{IacopiniZavattini}. The
electronic extinction given by the substraction of the electric
signals corresponding to the intensities of the two rays exiting
the analyzer prism play the role of the optical extinction
$\sigma^2$. The presence of a static uncompensated ellipticity
$\Gamma$ gives a $DC$ component to the signal, as any modulated
ellipticity gives a corresponding modulation to the signal
\cite{Rizzo}.

In practice all the results of the previous paragraph apply also
for the balanced polarimetry.

\subsection{Correlation between the signal and the magnetic pulse}

As we have shown in the previous paragraphs, the ellipticity can
be written as $\Psi(t)=\alpha B(t)^2$ or equivalently :

\begin{equation}\label{Psit}
\Psi(t)=\frac{I_e(t)-I_{DC}}{2I_t\Gamma},
\end{equation}
\noindent where $I_{DC}$ has been defined in sec. \ref{sec:1}, and
we assume that the average effect given in formula \ref{moy} can
be neglected.

The signal analysis based on the Fourier spectrum discussed in the
previous paragraph is the most appropriate only when the
ellipticity $\Psi(t)$ is a harmonic function of time. As we use a
pulsed field, we will use a different technique.

To recover the value of the constant $\alpha$ one can use the
following formula

\begin{equation}\label{alpha}
\alpha=\frac{\int_0^T\Psi(t)B(t)^2dt}{\int_0^TB(t)^4dt}
\end{equation}

\noindent with $B(0)=B(T)=0$.

An interesting property of formula \ref{alpha} is that if
$\Psi(t)$ is proportional to the derivative of $B(t)$ or $B(t)^2$,
$\alpha=0$. This means that this technique allows an easy
rejection of this kind of spurious signals.

On the other hand, formula \ref{alpha} gives a value of
$\alpha\neq 0$ even if $\Psi(t)$ is not proportional to $B(t)^2$.
A slightly more complicated formula can be used

\begin{equation}\label{alpha'}
\alpha'(\tau)=\frac{\int_0^T\Psi(t'+\tau)B(t')^2dt'}{\int_0^TB(t')^4dt'}
\end{equation}

It is evident that $\alpha=\alpha'(0)$. Now one can compare the
function $\alpha'(\tau)$ with

\begin{equation}\label{alpha'_B}
\alpha'_B(\tau)=\frac{\int_0^T
B(t'+\tau)^2B(t')^2dt'}{\int_0^TB(t')^4dt'}
\end{equation}

Only if $\Psi(t)$ is proportional to $B(t)^2$, $\alpha'(\tau)$ is
proportional to $\alpha'_B(\tau)$ for any value of $\tau$.

It is important to notice that $I_e(t)$ will usually be measured
at $\theta = 45^{\circ}$ which gives the maximum value for
$\Psi(t)$ (see formula \ref{Elli}). Since at $\theta = 0^{\circ}$,
$\Psi(t) = 0$, even if $B(t) \neq 0$, the measurement of $I_e(t)$
at $\theta = 0^{\circ}$ and the comparison of it with $I_{DC}$ is
a crucial test. Any difference from zero of the quantity
$I_e(t,\theta=0^{\circ}) - I_{DC}$ is an indication of noise
induced by the magnetic field on the apparatus during the pulse.

\section{Magnet and cryostat}
\label{sec:2}

A first analytical calculation of a model pulsed coil geometry has
been presented in ref. \cite{Aske}. A field as high as 25 T over a
meter length seemed to be achievable.  This model has been
considered as the starting point of the actual coil design in the
framework of the BMV project. \cite{BMV}. The basic idea was to
get the current creating the magnetic field as close as possible
of the light path over a length as long as possible. For the
magnetic birefringence application, one has to maximize the
integral of the square of the field over the magnet length L :

\begin{equation}\label{}
    B^2L=B_0^2L_{eq}=\int_{-L/2}^{L/2}B_y^2(z)dz
\end{equation}

\begin{figure}[h]
\includegraphics[width=9cm]{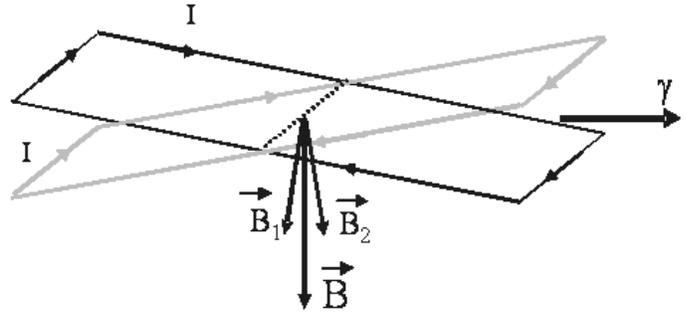}\\
  \caption{Scheme of the Xcoil.}\label{X}
\end{figure}

Here, if $B_0$ is the  field maximum, we define $L_{eq}$ as the
equivalent length of a magnet giving an uniform field value $B_0$
on the axis. The ratio $L_{eq}/L$ gives us a measure of the
field's uniformity. In fig. \ref{X} we show the configuration we
have studied and realized.

Because of the particular geometry, in the following we will call
it Xcoil \cite{Portugall}.

\begin{figure}[h]
  \includegraphics[width=9cm]{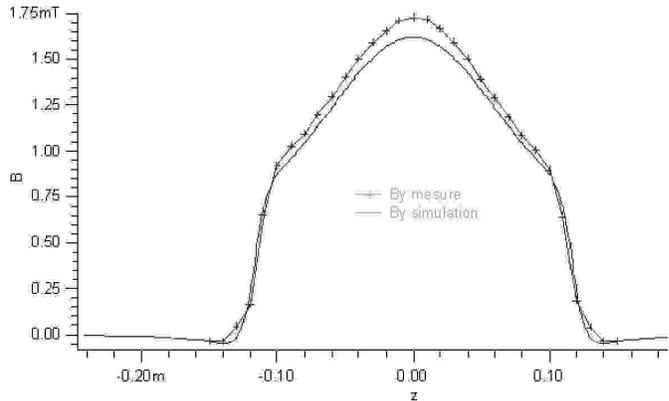}\\
  \caption{Computed (line) and measured (line and markers) field profile.}\label{Campi}
\end{figure}

In the approximation that the length of the coil is much bigger
that its width, one can calculate the field in the centers of the
Xcoil by using the expression of the field created by four
infinite wires. Following these analytical results, we have
realized an actual coil based on the Xcoil scheme. In fig.
\ref{Campi} we can see the comparison between experimental and
theoretical field computed by finite element analysis. Here, the
field is measured by a pick-up coil and a lock-in amplifier for an
oscillating current of 1 A at 230 Hz.

Our parameters leads to a field factor of 1.68 T/kA, an equivalent
length $L_{eq}$ of 13.2 cm : almost half of the total magnet
length. The relevant electrical parameters are L=400 $\mu H$ and
$R_{@77K}=52\;m\Omega$.

The higher experimental field in fig. \ref{Campi} is explained by
the shape of the wire used that does not exactly correspond to the
infinitesimal filaments in the simulation.

The coil support is manufactured using G10, a composite material
commonly used to deal with high stresses and cryogenic conditions.
External reinforcements from the same material are added after
winding to contain the magnetic pressure that can be as high as
500 MPa at the field maximum.

We have produced several Xcoils, having outer dimensions of 250 mm
x 100 x 100 mm. The internal hole available for optical
measurements is of 17 mm. This value is mainly given by the
dimension of the vacuum tube and thickness of the tubes for
double-walled cryostat.

Tests at high field have been performed using a pulsed power
supply, that consists of six capacitors switched by thyristors.
This capacitor bank is directly down-scaled from the LNCMP's 14 MJ
capacitor bank and it containes a maximum energy of 200 kJ.

\begin{figure}[h]
\begin{center}
  \includegraphics[width=4cm]{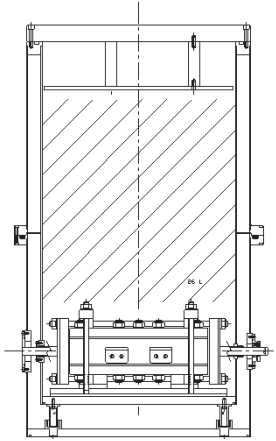}\\
  \caption{Drawing of the liquid nitrogen cryostat. The coil is also shown.}\label{Cryo}
  \end{center}
\end{figure}

Due to the increase of resistance by the Joule heating during the
shot, the pulse duration is voltage-dependant : it takes 18 ms at
low fields (4T) and 6 ms at 14 T. Like for conventional pulsed
magnets, the coil is placed in a liquid nitrogen cryostat to limit
heating consequences. The whole cryostat is double-walled with a
vacuum thermal insulation, including for the inner tube. As shown
in fig. \ref{Cryo}, the cryostat provides two additional
functions. It houses the cavity which passes through an opening at
room temperature arranged through the cryostat and it allows the
cavity to be mechanically disconnected from the coil. Dry and warm
nitrogen gas coming from the main bath is reinserted between the
cavity and the internal bore of the cryostat in order to prevent
any trapping of air moisture. Finally, the cryostat allows for
fast disassembly and replacement in the event of degradation of
the coil.

\begin{figure}[h]
  \includegraphics[scale=0.45]{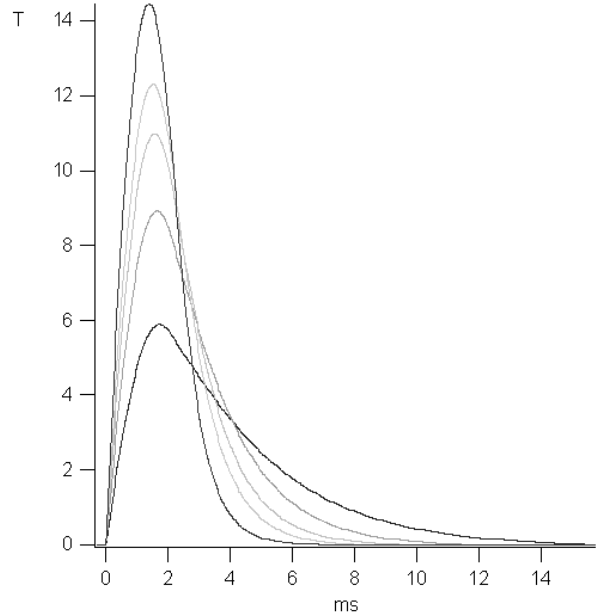}\\
  \caption{Series of magnet pulses.}\label{Tiri}
\end{figure}

In fig. \ref{Tiri} we show a series of magnet pulses obtained by
increasing the power supply voltage. The maximum field obtained
without generating significant resistance changes in the coil,
that would indicate the onset of conductor ageing, has been 14.3
T.

The long-term effect of aging at lower field has been studied by
pulsing a coil for 100 times at 11.5 T and 100 times at 12.5 T. No
significant change in its resistance has been detected.

\begin{figure}[h]
  \includegraphics[width=8cm]{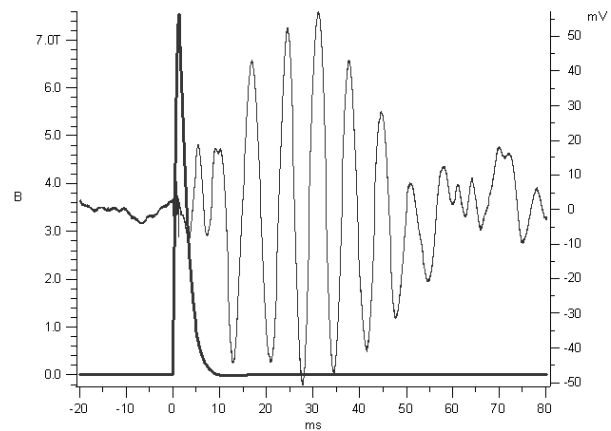}\\
  \caption{Mechanical noise during and after a pulse.}\label{Meca}
\end{figure}

Another crucial point for optical application is that the
mechanical noise created by the coil during the pulse has to be as
low as possible. We measured for one of our coils, the mechanical
noise on the floor of the test area during the pulse. The pulse
was 7.5 T high and its duration was about 5 ms. In fig. \ref{Meca}
we show such a measurement. An interesting feature is that the
shock wave arrives after the pulse. The mechanical path was less
than one meter.

During normal operation, by measuring the resistance after each
shot, a pick-up signal to monitor the field value and the
frequency in vibration spectrum, we can monitor the coil behavior.

\begin{figure}[h]
  \includegraphics[width=9cm]{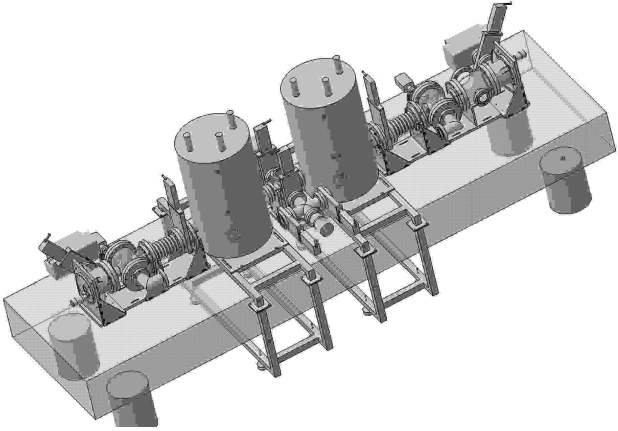}\\
  \caption{Drawing of our experimental set up with the two cryostats in place.}\label{Manip}
\end{figure}

Fig. \ref{Manip} shows a drawing of the experimental set up
corresponding to the Fabry-Perot cavity and the field region. Two
cryostats are shown. The ultra high vacuum chambers for the optics
sit on an optical table 3.6 meter long. The length of the cavity
is 2.2 meters. The cryostats are supported above the optical table
by a structure mechanically decoupled from the optical cavity and
the vibrations's path to the mirrors is more than 2 meters long.

\section{Vacuum system}
\label{sec:4}

\begin{figure}[h]
  \includegraphics[width=9cm]{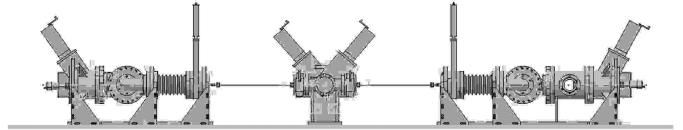}\\
  \caption{Scheme of our vacuum system.}\label{Vide}
\end{figure}

In fig. \ref{Vide} we show a scheme of our vacuum system. The most
critical parts are the two tubes that allow light to pass through
the magnetic field region. Their length is about 50 cm and the
inner diameter is 10 mm.

The main point is that the presence of residual gas in the vacuum
pipe induces an ellipticity on the light beam because of the
magnetic birefringence of gases (Cotton-Mouton effect)
\cite{RizzoRizzo}. To keep such an ellipticity small compared to
the one to be measured, the maximum pressure $P_g$ of a residual
gas constituent has to satisfy the following formula

\begin{equation}\label{CMgas}
P_g(atm) \ll \frac{\lambda \Psi(t)}{2F L \Delta n_u B(t)^2}
\end{equation}

where following \cite{RizzoRizzo} $\Delta n_u$ is the anisotropy
of the index of refraction for the residual gas component for a
magnetic field of 1 T and a pressure of 1 atm at a temperature of
0 $^\circ$C. For example, if $\Psi(t) \simeq 10^{-9}$, $F \simeq
5\times 10^{5}$, $L \simeq 1$ m, $\Delta n_u \simeq 10^{-12}$ as
for $O_2$ gas, $B(t) \simeq 10$ T, $P_g(atm) \ll 7.6\times10^{-9}$
torr.

The vacuum system set up allows a dry roughing by a spiral pump
and a turbo molecular pump, then a permanent vibration free
pumping thanks to two ion pumps. As tests have shown, in the
vacuum pipe a pressure better than 10$^{-8}$ mbar is expected,
while a slightly better pressure should be reached in the vacuum
chambers where the mirrors are inserted. Finally, a gas analyzer
will be put in between the two vacuum pipe passing through the
magnets to check the nature of the residual gas, and monitoring
the Cotton-Mouton of the residual gas.

\section{Magnetic shielding}
\label{sec:5}

The stray magnetic field of the magnet induces systematic effects
on the optics and especially on the mirrors which are the elements
closest to the magnets. Mirror Faraday effect, i.e. the rotation
of the polarization of a linearly polarized light induced by a
magnetic field perpendicular to the mirror surface, has been
reported in ref. \cite{Iacopini}. Faraday rotation was measured
being of the order of $K_F=3.7\times10^{-6}$ rad T$^{-1}$ per
reflection.  Mirror Cotton-Mouton effect, i.e. the ellipticity
induced on a linearly polarized light by a magnetic field parallel
to the mirror surface, has been reported in ref.
\cite{Bialolenker} and it amounted to about $K_{CM}=10^{-9}$
T$^{-2}$ per reflection. The mirrors tested in ref.
\cite{Bialolenker} were of different quality than the ones used in
ref. \cite{Iacopini}. Both results depend on the number of
reflecting layers and on the layer materials. However they can be
used to estimate the maximum stray field tolerable at the mirror
location. For example, the ellipticity induced by the mirrors
because of their own Cotton-Mouton effect is negligible compared
to the ellipticity to be measured, when the stray magnetic field
parallel to the mirror surface obeys the following relation.

\begin{equation}\label{StrayField}
B_{stray}^\parallel\ll\sqrt{\frac{\pi}{2F}\frac{\Psi_0}{K_{CM}}}
\end{equation}

For example, if $\Psi_0 \simeq 10^{-9}$ and $F \simeq 500 000$,
$B_{stray}^\parallel \ll 2\times 10^{-3}$ T.

An equivalent formula can be found for the component of the stray
magnetic field perpendicular of the mirror surface.

\begin{equation}\label{StrayField}
B_{stray}^\perp\ll\frac{\pi}{2F}\frac{\rho_0}{K_{F}}
\end{equation}

where $\rho_0$ is the rotation in which one would be interested,
and we have assumed as usual that the number of reflections on the
mirrors in a Fabry-Perot cavity is equal to ${2F}/{\pi}$.

Using one of our Xcoils, powered for a central field of 7 T, we
have measured the stray field on the axis of the coil. At 70 cm
from the coil's center the field is of the order of 1 mT$/$kA,
which corresponds to a reduction of a factor 500 with respect to
the field at the centre of the magnet. We have also measured that
the shielding factor given by a 4 mm copper plate is almost 5 at
this field. In our experiment, the current is around 8000 A, what
gives us a field on the nearest mirror of 0,1 mT. With a shielding
of at least 80\% (we certainly can expect more than 90 \%), it
gives a field on the mirror of 0,02 mT. If we need a better
shielding, we only need to put a second copper plate. This is a
major advantage of pulsed fields over static fields.

\section{Current status and perspectives}
\label{sec:7}

The whole optical system is operational, and the laser has been
locked to different cavities. In particular, the laser has been
locked to linear cavities (up to 2 m length) and ring cavities
(round trip of 2.4 m) of finesses up to 50 000. The measuring
technique has been tested by measuring a birefringence induced by
an electric field in a gas (Kerr effect) of about 10$^{-16}$
\cite{Bielsa}.

After years of developments and tests, we have finally put optics
and magnet together and started debugging and studying
sensitivity. First with just one magnet in place, then, when test
runs will be completed, with two magnets in place. This
configuration corresponds to 40 T$^2$m. Finesse greater than 300
000 is expected. Sensitivity should also be at least $10^{-8}$ rad
Hz$^{-1/2}$, thanks to the high central frequency of the modulated
effect. All these experimental parameters exceed the corresponding
values for the PVLAS experiment. Therefore, we will be able to
test the PVLAS results using this first version of our apparatus.

Once this first step is accomplished, we will continue towards the
QED vacuum magnetic birefringence measurement. The critical points
will be to reduce laser noise to reach a sensitivity very close to
the quantum limit and to continue the magnet R\&D to reach 25 T. A
3-D complete computer modelling of our coils will be implemented
to study the behavior of our coils under constraints.

In any case, we will need more powerful transverse pulsed magnets.
Xcoils with a field region of 25 cm have been successfully tested,
we are confident that final apparatus will consist of magnets
capable to deliver a field over a length such that $B_0^2 L = 600$
T$^2$m. QED ellipticity to be measured will be of the order of
$4\times10^{-9}$ rad, which should be reached in few hundreds of
magnet pulses corresponding to a few weeks of data acquisition.

\section{Acknowledgements}

We thank J. Billette, M. Fouch\'e, D. Forest, J-P. Laurent, J.
Mauchain, J-L. Montorio, L. Polizzi, and A. Zitouni. We also
acknowledge the strong support of the engineering and technical
staff of LCAR, LNCMP and LMA. This work is supported by the {\it
ANR-Programme non th\'{e}matique} (ANR-BLAN06-3-139634), and by
the {\it CNRS-Programme National Astroparticules}.

\end{document}